\documentclass[aps,floatfix,onecolumn,a4paper,showpacs, nofootinbib,superscriptaddress,11pt]{revtex4}
\usepackage{graphicx,float}\usepackage{graphicx,float}
\usepackage[all]{xy}
\usepackage{amsmath,upgreek}
\usepackage{amssymb}
\usepackage{color}
\usepackage{epsfig,bm}		
\usepackage{graphicx,epstopdf}
\usepackage{subfigure}
\usepackage{pdfpages,slashed}
\usepackage[colorlinks,hyperindex]{hyperref}

\definecolor{green1}{RGB}{0,128,0}
\hypersetup{backref=true,pagebackref=true}
\hypersetup{%
  colorlinks = true,
  linkcolor  = blue,
  citecolor = cyan,
}
\usepackage{bookmark,textgreek}
\usepackage{hyperref,color,xcolor}
\hypersetup{hidelinks,hyperindex=true,colorlinks=true,breaklinks=true,urlcolor= blue}
\hypersetup{%
  colorlinks = true,
  linkcolor  = blue
}\usepackage{amssymb}
\newsavebox{\foobox}

\usepackage{graphicx,float,tikz}
\usepackage[all]{xy}
\newcommand\ringring[1]{%
  {
   \mathop{\kern0pt #1}\limits^{
     \vbox to-1.85ex{
       \kern-2ex 
       \hbox to 0pt{\hss\normalfont\kern.1em \r{}\kern-.45em \r{}\hss}%
       \vss 
     }
   }
  }
}\newcommand\orcidroldao{{\href{https://orcid.org/0000-0003-3978-532X}{\orcidicon}}}
\newcommand{\orcidicon}{%
	\begin{tikzpicture}
	\draw[lime, fill=lime] (0,0)
		circle [radius=0.16]
		node[white] {{\fontfamily{qag}\selectfont \tiny ID}};
	\draw[white, fill=white] (-0.0625,0.095)
		circle [radius=0.007];
	\end{tikzpicture}	\hspace{-2mm}
}
\newcommand{\bpartial}{\mathop{\partial\kern -4pt\raisebox{.8pt}{$|$}}}

\newcommand{\bes}{\begin{subequations}}
\newcommand{\ees}{\end{subequations}}
\def\beq{\begin{eqnarray}}
 \newcommand{\bltx}{\textcolor{black}}

    \newcommand{\ptx}{\textcolor{black}}
\def\eeq{\end{eqnarray}}
\def\be{\begin{equation}}
\def\ee{\end{equation}}
\newcommand{\n}{\nonumber\\}
\newcommand{\non}{\nonumber}
\newcommand{\kerr}{a_{\textrm{\tiny{Kerr}}}}

\usepackage{cancel}
\usepackage{textcomp}

\begin{document}

\title{Generalized Navier--Stokes equations and soft hairy horizons in fluid/gravity correspondence}

\author{A. J. Ferreira--Martins}
\email{andre.juan@aluno.ufabc.edu.br}
\affiliation{Federal University of ABC, Center of Natural Sciences, Santo Andr\'e, 09580-210, Brazil}

\author{R. da Rocha\orcidroldao\!\!}
\email{roldao.rocha@ufabc.edu.br}
\affiliation{Federal University of ABC, Center of Mathematics, Santo Andr\'e, 09580-210, Brazil}


\begin{abstract}
The fluid/gravity correspondence establishes how gravitational dynamics, as dictated by Einstein's field equations, are related to the fluid dynamics, governed by the relativistic Navier--Stokes equations. In this work the correspondence is extended, where the duality between incompressible fluids and gravitational backgrounds with soft hair excitations is implemented. This construction is set through appropriate boundary conditions to the gravitational background, leading to a correspondence between generalized incompressible Navier--Stokes equations and soft hairy horizons.
\end{abstract}



\maketitle

\section{Introduction}

AdS/CFT is well known to assert the duality between $\mathcal{N}=4$ SU($N$) super Yang-Mills theory and type
IIB string theory with AdS${}_5 \times S^5$ compactification \cite{malda}. In this correspondence, gravitons in the AdS${}_5$ bulk can be mapped to the energy-momentum tensor governing the strongly coupled theory of the AdS${}_5$ boundary theory. In the low energy limit, hydrodynamics on the boundary can be ruled by the conservation of the
stress tensor, implementing a sector of
universal decoupled energy-momentum tensor dynamics.
In the long
wavelength regime, the conservation of the energy-momentum tensor is led to hydrodynamics, as an explicit correspondence between gravity and fluid dynamics. It consists of the fluid/gravity correspondence, where solutions of relativistic hydrodynamics on the boundary yield solutions of Einstein's field equations, also describing black holes 
\cite{hubeny_hard,Haack:2008cp,Policastro:2002se}. 
Hydrodynamics equations can then lead to the Navier--Stokes 
equations describing incompressible and viscous fluid dynamics. Conformal symmetries at the gravity sector were shown to be dual to accelerated boost symmetries of the Navier--Stokes equations \cite{Bhattacharyya:2008kq}. Besides, fluid/gravity correspondence and the membrane paradigm are closely linked \cite{Pinzani-Fokeeva:2014cka}. 

One can implement the fluid/gravity correspondence using isometries to perturb a black brane solution of Einstein's field equations, describing gravity in AdS${}_5$. 
\bltx{After promoting the parameters of the boost isometries to local functions, the transformed solution does not satisfy Einstein's field equations, in general.} Equations of motion, except the local parameters, satisfy  
Navier--Stokes regarding the field theory on the boundary. 
One of the most impressive results after fluid/gravity is the shear viscosity-to-entropy density ratio \cite{kss}. Initially, the lower bound $\hbar/4\pi k_B$, where $k_B$ denotes the Boltzmann constant, was derived for strongly-coupled QFTs that are dual theories to gravity and black branes in AdS${}_5$. Thereafter, deformed AdS${}_5$--Schwarzschild black branes were derived, with extended shear viscosity-to-entropy density ratios \cite{Ferreira-Martins:2019svk,Ferreira-Martins:2019wym,Casadio:2016zhu}.  
Fluid/gravity was also employed to study compact distributions \cite{glueball,daRocha:2020jdj}. \bltx{Relativistic Navier--Stokes equations were shown to be stable and causal, with suitable definitions of hydrodynamic variables outside of equilibrium \cite{Hoult:2020eho}}.

Translational symmetries yield conservation laws in theories describing gravity when the backgrounds are asymptotically flat \cite{daniel3}. These laws require black holes to carry a large amount of soft 
hair. Soft hair was introduced for soft bosonic fields on the black hole horizon, demonstrating how information regarding their quantum state is compressed into holographic loci at the future boundary associated with 
the black hole horizon \cite{Hawking:2016msc}. 
Soft hair was also shown to drive final states of the black hole evaporation process \cite{Mirbabayi:2016axw}, previously precluded by charge and energy conservation laws. Gravitational decoupling methods have been also used to study black hole primary hairs, circumventing the no-hair theorem \cite{Ovalle:2020kpd}.

\bltx{There exists a correspondence between vacuum Einstein's field equations solutions in $(D+2)$ dimensions and incompressible Navier--Stokes equations in $(D+1)$ dimensions, which implements a holographic duality between the theories \cite{minwalla}. In this work, we establish this result in the case where the gravitational background has soft hair excitations. } Besides, as it was shown in Ref. \cite{fg2}, the near-horizon expansion in gravity is formally identical to the hydrodynamic expansion in fluid dynamics, which then allows us to precisely state that horizons are incompressible fluids. Emulating the results of Refs. \cite{fg2, fg3}, the chosen boundary conditions fix a flat induced metric on a cutoff manifold, corresponding to a radial foliation $\Sigma_c$. Afterward, we generalize the boundary conditions, allowing the induced metric on $\Sigma_c$ to fluctuate. These boundary conditions lead to soft hair excitations in the black hole horizons \cite{sh}, which was proved to hold in arbitrary dimensions \cite{daniel3}. \bltx{Besides the influential developments in Refs. \cite{Donnay:2019jiz,Donnay:2015abr,Donnay:2016ejv,Donnay:2020guq}, with these new boundary conditions we study a possible generalization of the incompressible Navier--Stokes equations describing the dual fluid at a soft hairy horizon.}
\bltx{The induced metric has the Bondi--van der Burg--Metzner--Sachs (BMS)-like symmetry group, which complies with the black hole membrane paradigm, where black hole event horizons behave like fluid
membranes \cite{Penna:2015gza}. BMS-like supertranslations and superrotations have been already explored in the study of the geometry of black hole horizons, described as a Carrollian geometry emerging from an ultra-relativistic limit \cite{Donnay:2019jiz}. }

This paper is organized as follows: in Sec. \ref{sec:hydro} the hydrodynamic limit of fluid dynamics is briefly reviewed, through the invariance of solutions to the incompressible Navier--Stokes equations under the hydrodynamic scaling. In Sec. \ref{sec:setup}, the general gravitational scenario which will be considered in this work is introduced. Thereafter, Sec. \ref{sec:fixed} is devoted to emulating the developments of \cite{fg1, fg2}, establishing a particular fluid/gravity correspondence, given the imposition of suitable boundary conditions to the gravitational scenario. In Sec. \ref{sec:varying} we then generalize these boundary conditions, moving to ones which were shown to lead to soft hair excitations \cite{daniel3}, ultimately leading to a generalization of the original fluid/gravity correspondence. Sec. \ref{sec:conc} is dedicated to conclusions, remarks, and outlook.

\section{The hydrodynamic limit of fluid dynamics}
\label{sec:hydro}

For interacting QFTs, the fluid/gravity correspondence holds for near-equilibrium dynamics, in the long-wavelength regime.
The hydrodynamic limit requires looking at the system under scrutiny at a length scale that is large enough when contrasted to the mean free path length, which comprises the inherent length scale of the system. It prevents any microscale inhomogeneity \cite{Bhattacharyya:2008kq} to set in.  
The relativistic Navier--Stokes equations can be derived using a gravitational prescription, avowing fluid dynamics to be realized as the long-wavelength effective portrayal of conformal field theories that are strongly coupled. 
In component notation, the incompressible Navier--Stokes equations in $D$ spatial dimensions read
\begin{subequations}
\beq\label{1a}
\partial^i v_i &=& 0,\\
\partial_t v_i - \eta \partial^2 v_i + \partial_i p + v^j \partial_j v_i &=& 0,\label{1b}
\eeq
\end{subequations}
\noindent where $i, j, k = 1,\ldots, D$; $\eta$ is the kinematic viscosity, 
$v_i = v_i(t, x^k)$ denotes the velocity field, and $p = p(t, x^k)$ is the pressure field. These equations are solved for an incompressible fluid described by the pair $(v_i, p)$. A non-relativistic rescaling, as well as a rescaling of the amplitude of these solution fields under the parameter $\epsilon$, consist of 
\begin{subequations}
\beq\label{nn_1}
v_i^{(\epsilon)}(t, x^i) = \epsilon v_i(\epsilon^2 t, \epsilon x^i) \equiv \bltx{\mathring{v_i}}, 
\\
p^{(\epsilon)}(t, x^i) = \epsilon^2 p(\epsilon^2 t, \epsilon x^i) \equiv \bltx{\mathring{p}}.\label{nn_2}
\eeq
\end{subequations}
\textcolor{black}{For the remaining of this manuscript, we will drop the superscript $(\epsilon)$, and define $v_i^{(\epsilon)} \equiv \bltx{\mathring{v_i}}$; $p^{(\epsilon)} \equiv \bltx{\mathring{p}}$, as the $\epsilon$-parametrized fields. This will facilitate the notation throughout the text.}

The transformations (\ref{nn_1}, \ref{nn_2}) are hydrodynamic scalings, explicitly amounting to the mappings
\beq 
\label{3b}
\left\{\begin{matrix}
x^i \mapsto x^i = \epsilon x^i\\ 
\partial_i \mapsto \partial_i = \epsilon \partial_i
\end{matrix}\right.,\;\;\;\;\; 
\left\{\begin{matrix}
t \mapsto t = \epsilon^2 t \\ 
\partial_t \mapsto \partial_t = \epsilon^2 \partial_t
\end{matrix}\right..
\label{eq:scale}
\eeq
\noindent They yield the incompressible Navier--Stokes equations (\ref{1a}, \ref{1b}) to be respectively mapped to
\begin{subequations}
\beq
\partial^i \bltx{\mathring{v_i}} = 0,\\
\partial_t \bltx{\mathring{v_i}} - \eta \partial^2 \bltx{\mathring{v_i}} + \partial_i \bltx{\mathring{p}} + \bltx{\mathring{v}^j} \partial_j \bltx{\mathring{v_i}} = 0.
\eeq
\end{subequations}
Therefore, the pair $(\bltx{\mathring{v_i}}, \bltx{\mathring{p}})$ is a family of solutions of the incompressible Navier--Stokes equations, built 
\textcolor{black}{from}
the original solution pair $(v_i, p)$ under a parameterization by $\epsilon$, according to Eqs. (\ref{nn_1}, \ref{nn_2}).

The hydrodynamic limit is achieved under the hydrodynamic $\epsilon$-scaling when $\epsilon \rightarrow 0$. In this case, the higher-derivative corrections to the Navier--Stokes equations that may appear for real fluids become irrelevant, since they scale at order $\mathcal{O}(\epsilon^2)$. For this reason, the incompressible Navier--Stokes equations are adequate to describe any fluid in the hydrodynamic limit\textcolor{black}{, given that, in this limit, higher-order dissipative corrections vanish}.

Notice that the hydrodynamic scaling summarized in Eq. \eqref{eq:scale} is such that
\beq
 \bltx{\mathring{v_i}} &\sim \mathcal{O}(\epsilon),\qquad \bltx{\mathring{p}} \sim \mathcal{O}(\epsilon^2),\qquad 
 \partial_i \sim \mathcal{O}(\epsilon),\qquad \partial_t \sim \mathcal{O}(\epsilon^2).
 \label{eq:hydro_scale}
\eeq
These are important relations to be used when the hydrodynamic expansion 
\textcolor{black}{of} the gravitational background will be employed.

\section{The general setup}
\label{sec:setup}

The general gravitational setup is implemented considering geometries in $(D+2)$ dimensions, covered by the chart $\{t, r, x^i\}$ and with an outer boundary. The cutoff codimension one manifold, $\Sigma_c$, is defined by the radial foliation
$ \Sigma_c: r-r_c = 0$, with coordinates $x^a = \{t, x^i\}$. In what follows, the unit normal covector  field on $\Sigma_c$ reads $
n_{\mu}\mathrm{d}x^{\mu} = \left({g^{rr}}\right)^{-1/2} \mathrm{d}r,$
 whereas the corresponding normal vector is given by $
n^{\mu}\partial_{\mu} = \left({g^{rr}}\right)^{-1/2} \left ( g^{rt} \partial_t + g^{rr}\partial_r + g^{ri} \partial_i \right)$. 
The induced metric $\gamma_{\mu \nu} = g_{\mu \nu} - n_{\mu}n_{\nu}$ 
 on $\Sigma_c$ yields the extrinsic curvature 
\begin{equation}
\begin{aligned}
 K_{\mu \nu} =
\frac{1}{2} \left ( n^{\sigma} \partial_{\sigma} \gamma_{\mu \nu} + \gamma_{\sigma \nu}\partial_{\mu}n^{\sigma} + \gamma_{\mu \sigma} \partial_{\nu} n^{\sigma} \right ) \ .
 \label{eq:ext_curv}
\end{aligned}
\end{equation}
The Brown--York stress tensor on $\Sigma_c$ is then defined in terms of the extrinsic curvature as \cite{Brown:1992br}
\begin{equation}
 T_{a b}^{\scalebox{.55}{BY}} = 2 \left ( \gamma_{a b} K - K_{a b} \right ) \ ,
 \label{eq:brown_york}
\end{equation}
\noindent where $K = \gamma^{a b} K_{a b}$, 
and units in which $G = 1/16\pi$ are used. 
Imposing conservation of the Brown--York stress tensor, it can be put into the form of a perfect fluid satisfying the incompressible Navier--Stokes equations \cite{fg2}. Therefore, one can reduce Einstein's field equations to the incompressible Navier--Stokes equations, as will be subsequently detailed. Two different boundary conditions for the induced metric on $\Sigma_c$ can be obtained. First keeping it fixed, and then allowing it to fluctuate. Either choice, which corresponds to a boundary condition imposed to the $(D+2)$-dimensional geometry, leads to relevant yet different results in the fluid/gravity correspondence context, as will be constructed in the next section.

\ptx{There is an important remark concerning the relativistic nature of the fluids under consideration. In what follows, the bulk solution is constructed under the non-relativistic, hydrodynamic $\epsilon$-expansion of the properties of the fluids. Eqs. (\ref{1a}, \ref{1b}), which we will be the seed of the procedure leading to the dual fluid, manifestly describe a non-relativistic incompressible fluid. Thus, it is reasonable to question whether the fluid dual to Einstein gravity on $\Sigma_c$ is, in fact, a relativistic fluid. As stated by Ref. \cite{Bhattacharyya:2008kq}, Eqs. (\ref{1a}, \ref{1b}) are the result of a specific scaling limit applied to any reasonable relativistic fluid with a hydrodynamic equation of state. Here, this limit regards long distances, long times, and low amplitudes, as detailed in Sec. \ref{sec:hydro}. In this limit, the equations become non-relativistic and incompressible, as the velocities under consideration are much lower than the speed of light in this medium \cite{landau_fluids}. Ref. \cite{Bhattacharyya:2008kq} shows how the Navier--Stokes equations are derived via the application of appropriate scalings to the equations of relativistic hydrodynamics, with an arbitrary equation of state, although conformal symmetry is a particular case addressed in detail in that work. Under this scaling, the equation $\nabla_\mu T^{\mu \nu}=0$ does reduce to Eqs. (\ref{1a}, \ref{1b}). Therefore, a non-relativistic description of the fluid is indeed adequate, even though the fluid in the dual theory is relativistic, as long as we work with fluid variables under the hydrodynamic expansion, and consider the hydrodynamic limit.}

\ptx{Besides, Ref. \cite{fg3} provides a full description of the fluid, which is obtained from a hydrodynamic relativistic stress tensor, even though the procedure leading to the duality, which employs the hydrodynamic expansion, is implemented in terms of the non-relativistic incompressible Navier--Stokes equations, for the aforementioned reasons. In such a description, which includes dissipative terms, the authors elaborate on the interpretation and implications of some unusual features of the dual fluid, as its zero energy density but nonzero pressure in equilibrium. Later on, we shall further comment on these properties and their implications.}

\section{A fixed induced metric}
\label{sec:fixed}

A boundary condition for the $(D+2)$-dimensional spacetime is considered by a fixed induced metric on $\Sigma_c$, given by
\begin{equation}
\gamma_{a b} \mathrm{d}x^a \mathrm{d}x^b = -r_c \mathrm{d} t^2 + \delta_{ij}\mathrm{d}x^i \mathrm{d}x^j, 
\label{eq:bc1_down}
\end{equation}
\noindent where $\delta_{ij}$ is the $D$-dimensional boundary metric of $\Sigma_c$, represented by the Kronecker delta components \cite{fg2, fg3}. It corresponds to fixing a flat induced metric on $\Sigma_c$, since $\gamma_{ab} = \eta_{ab}$ corresponds to the Minkowski metric when $\sqrt{r_c}$ is regarded as the speed of light.

A $(D+2)$-dimensional geometry which is straightforwardly compatible to this boundary condition is the Rindler spacetime,
\begin{equation}
 g_{\mu \nu} \mathrm{d}x^\mu \mathrm{d}x^\nu = -r \mathrm{d}t^2 + 2 \mathrm{d}t \mathrm{d}r + \delta_{ij}\mathrm{d}x^i \mathrm{d}x^j \ , 
 \label{eq:rindler_st}
\end{equation}
\noindent with ingoing Rindler coordinates $\{r, t, x^i\}$. Notice that setting $t = 2 \ln (X + T)$ and $4r = X^2 - T^2$ yields $g_{\mu \nu} \mathrm{d}x^\mu \mathrm{d}x^\nu = -\mathrm{d}T^2 + \mathrm{d}X^2 + \delta_{ij}\mathrm{d}x^i \mathrm{d}x^j$, corresponding to flat spacetime. The boundary condition is satisfied at $r=r_c$.

The induced metric reads 
\begin{equation}
 \gamma_{\mu \nu} \mathrm{d}x^{\mu} \mathrm{d}x^{\nu} = -r \mathrm{d}t^2 - \frac{1}{r} \mathrm{d}r^2 + 2 \mathrm{d}t \mathrm{d}r + \delta_{ij}\mathrm{d}x^i \mathrm{d}x^j, 
\end{equation}
\noindent which clearly reduces to Eq. \eqref{eq:bc1_down} at $r = r_c$. 
Using these results, the extrinsic curvature is calculated according to Eq. \eqref{eq:ext_curv}, which yields, on $\Sigma_c$,
\begin{equation}
 K_{ab}\mathrm{d}x^a\mathrm{d}x^b = -\frac{\sqrt{r_c}}{2} \mathrm{d}t^2,
\end{equation}
\noindent whose trace is given by $K = \frac{1}{2\sqrt{r_c}}$. Eq. \eqref{eq:brown_york} then implies the Brown--York stress tensor
\begin{equation}
\begin{aligned}
T_{a b}^{\scalebox{.55}{BY}} \mathrm{d}x^a \mathrm{d}x^b &= \frac{1}{\sqrt{r_c}} \delta_{ij} \mathrm{d}x^i \mathrm{d}x^j.
\end{aligned}
\end{equation}

Consider now the stress tensor of a perfect fluid, 
\begin{equation}
 T_{ab}^{\scalebox{.55}{PF}} = \rho u_a u_b + p \left (\gamma_{ab} + u_a u_b \right ), 
 \label{eq:perfect_fluid}
\end{equation}
\noindent where $\rho$ and $p$ are respectively the energy density and the pressure of the fluid in the local rest frame, whereas $u_a$ is the normalized fluid velocity. Notice that the choice $\rho = 0$, $p = r_c^{-1/2}$, $u_t = r_c^{1/2}$, and $u_i = 0$ yields
\begin{equation}\label{epf}
\begin{gathered}
T_{a b}^{\scalebox{.55}{PF}} \mathrm{d}x^a \mathrm{d}x^b = \frac{1}{\sqrt{r_c}} \delta_{ij} \mathrm{d}x^i \mathrm{d}x^j \ .
\end{gathered}
\end{equation}
\bltx{It is worth to mention that, in particular, the energy density $\rho$ vanishes, as there is no flat solution continuously connected to Rindler spacetime which generates a non-zero equilibrium energy density that concomitantly preservs the form of the induced metric on $\Sigma_c$ \cite{fg3}. Besides, one can formally define a trivial constant shift in the background energy density and pressure given by $\rho \mapsto \rho-C$ and $p \mapsto p+C$, for some constant $C$. This choice yields a constant density viscous fluid, what is physically more feasible than a zero density fluid, in what concerns the hydrodynamical fluid interpretation. This shift reflects into the inherent redundancy in the definition of the Brown--York stress tensor, $T_{ab}^{\scalebox{.55}{BY}}\mapsto T_{ab}^{\scalebox{.55}{BY}}+C\gamma_{ab}$ \cite{Brown:1992br}. Since it has no dynamical consequences, one can consider a viscous fluid with either non-null constant density or zero density. For simplicity, as formally equivalent to the constant density fluid flow case, Ref. \cite{fg3} suppresses it, setting $C=0$.}
Therefore, 
\begin{equation}
\begin{aligned}
&T_{a b}^{\scalebox{.55}{BY}} = T_{ab}^{\scalebox{.55}{PF}}, 
\label{eq:by=pf_1}
\end{aligned}
\end{equation}
\noindent and the Brown--York stress tensor is correspondent to that of a perfect fluid on $\Sigma_c$, with the fluid parameters as specified. This is our first hint towards fluid/gravity.

\subsection{Boosting the metric}
\label{sec:boost1}

\bltx{Two diffeomorphisms on the Rindler spacetime \eqref{eq:rindler_st} play an important role on the fluid/gravity correspondence and were studied in Refs. \cite{minwalla,fg2,fg3}. In particular, Ref. \cite{fg3} showed that  there are only two infinitesimal diffeomorphisms yielding metrics with the following properties: a) they admit a codimension one manifold $\Sigma_c$ of flat induced metric (\ref{eq:bc1_down}); b) the Brown--York stress tensor on $\Sigma_c$, given by \eqref{eq:brown_york}, takes the form of a perfect fluid stress tensor \eqref{eq:perfect_fluid}, as shown and displayed in Eq. \eqref{eq:by=pf_1}; c) they are stationary with respect to $\partial_t$ and homogeneous in the $x^i$ directions. Exponentiating these, the two finite diffeomorphisms -- respectively given by 
(\ref{two1}, \ref{two2}) and (\ref{two3}, \ref{two4}) -- are obtained}. The first diffeomorphism is a constant shift of the radial coordinate, replacing the Rindler horizon from $r=0$ to $r=r_h$, as well as a constant scaling of the temporal coordinate,
\bes
\beq\label{two1}
r &\mapsto& r - r_h,\\
t &\mapsto& \alpha t, \qquad \qquad 
\alpha = \left (1 - \dfrac{r_h}{r_c} \right )^{-1/2}.\label{two2}
\eeq\ees
The Rindler spacetime metric is accordingly transformed to
\begin{equation}
g_{\mu \nu} \mathrm{d}x^\mu \mathrm{d}x^\nu = -\alpha^2 (r-r_h) \mathrm{d}t^2 + 2\alpha \mathrm{d}t\mathrm{d}r + \delta_{ij}\mathrm{d}x^i \mathrm{d}x^j.
\end{equation}
The second diffeomorphism is a constant boost, 
\bes
\beq\label{two3}
t &\mapsto& \tilde{t} = \gamma \left (t - \dfrac{v_i x^i}{r_c} \right),
\\
x^i &\mapsto& \tilde{x}^i = x^i - \gamma t v_i + \left ( \gamma - 1 \right ) \dfrac{v_i v_j}{v^2} x^j,\label{two4}
\eeq
\ees
\noindent where $\gamma = \left (1 - {v^2}/{r_c} \right )^{-1/2}$. One can promptly calculate the transformed metric components. 
To keep the same sign convention in the components $\tilde{g}_{ri}$ and $\tilde{g}_{ti}$ as that of Refs. \cite{fg2, fg3}, the parity transformation $\tilde{x}^i \mapsto -\tilde{x}^i$ must be also applied. After dropping the tildes for simplicity, 
the following metric components transformed under the diffeomorphisms are obtained: 
\bes\beq\label{eq:boosted_metric_comps1}
\!\!\!\!\!\!\!\!g_{tt} &=& \gamma^2 \left ( v^2 - \alpha^2 \left ( r - r_h \right ) \right ), \qquad  
g_{tr} = \gamma \alpha,\qquad g_{ti} = \frac{\gamma^2 \alpha^2}{r_c} \left ( r - r_c \right ) v_i,\\
\!\!\!\!\!\!\!\!
g_{rr} &=& 0,\qquad g_{ri} = -\frac{\gamma \alpha}{r_c}v_i,\qquad 
g_{ij} = \delta_{ij} - \frac{\gamma^2 \alpha^2}{r_c^2} \left (r -r_c \right ) v_i v_j.
\label{eq:boosted_metric_comps}
\eeq\ees
%

Our goal now is to calculate the Brown--York stress tensor for the boosted metric, as already accomplished for the metric in ingoing Rindler coordinates. After calculating the inverse metric, the normal vector, and the induced metric on $\Sigma_c$, one arrives at the extrinsic curvature
\begin{equation}
\begin{aligned}
K_{ab}\mathrm{d}x^a \mathrm{d}x^b &= \gamma^2 \alpha \frac{\sqrt{r_c}}{2} \left ( \frac{2}{r_c}v_i \mathrm{d}x^i \mathrm{d}t \right) - \gamma^2 \alpha \frac{\sqrt{r_c}}{2} \left ( \mathrm{d}t^2 + \frac{1}{r_c^2} v_i v_j \mathrm{d}x^i \mathrm{d}x^j \right ), 
\end{aligned}
\end{equation}
\noindent whose trace $K = \frac{\alpha}{2\sqrt{r_c}}$ is employed in the calculation of the Brown--York stress tensor, according to Eq. \eqref{eq:brown_york},
\begin{equation}
\begin{aligned}
T_{a b}^{\scalebox{.55}{BY}} \mathrm{d}x^a \mathrm{d}x^b &= 
\frac{\alpha}{\sqrt{r_c}} \left ( \gamma^2 v^2 \mathrm{d}t^2 
-2\gamma^2 v_i \mathrm{d}x^i \mathrm{d}t \right )+ \frac{\alpha}{\sqrt{r_c}}
 \left ( \delta_{ij} + \frac{\gamma^2}{r_c}v_i v_j \right ) \mathrm{d}x^i \mathrm{d}x^j.
\end{aligned}
\end{equation}
Considering a perfect fluid characterized by {$\rho = 0$}, $ p = \alpha r_c^{-1/2}$, and $u_a = \gamma \left ( -r_c^{1/2} , r_c^{-1/2} v_i \right )$, its stress tensor, according to Eq. \eqref{eq:perfect_fluid}, reads 
\begin{equation}
\begin{aligned}
T_{a b}^{\scalebox{.55}{PF}} \mathrm{d}x^a \mathrm{d}x^b &= 
\frac{\alpha}{\sqrt{r_c}} \left ( \gamma^2 v^2 \mathrm{d}t^2 
-2\gamma^2 v_i \mathrm{d}x^i \mathrm{d}t \right )+ \frac{\alpha}{\sqrt{r_c}} \left ( \delta_{ij} + \frac{\gamma^2}{r_c}v_i v_j \right ) \mathrm{d}x^i \mathrm{d}x^j \ .
\end{aligned}
\end{equation}
Therefore, $
T_{a b}^{\scalebox{.55}{BY}} = T_{ab}^{\scalebox{.55}{PF}}$. 
The same correspondence between the Brown--York stress tensor and that of a perfect fluid on $\Sigma_c$ holds, as in Eq. \eqref{eq:by=pf_1}, although the transformations performed change the fluid properties. 
Its pressure has been rescaled by $\alpha$ and the spatial components appear on its velocity, which can be seen as a direct consequence of the boost. 


It is possible to explicitly introduce the fluid pressure as a parameter of the metric components, by substituting into Eqs. (\ref{eq:boosted_metric_comps1}, \ref{eq:boosted_metric_comps})  the relation $\alpha = p \sqrt{r_c}$, which yields 
\bes
\beq
\!\!\!\!\!\!\!\!g_{tt} &=& \gamma^2 \left ( v^2 - p^2 r_c \left ( r - r_h \right ) \right ), \quad\qquad 
g_{tr} = \gamma p \sqrt{r_c}, \quad\qquad g_{ti} = \gamma^2 p^2 \left ( r - r_c \right ) v_i,\label{eq:boosted_metric_presure_comps1}
\\  
\!\!\!\!\!\!\!\!g_{rr} &=& 0, \quad\qquad g_{ri} = -\frac{\gamma p}{\sqrt{r_c}}v_i, \quad\qquad  
g_{ij} = \delta_{ij} - \frac{\gamma^2 p^2}{r_c} \left (r -r_c \right ) v_i v_j.
\label{eq:boosted_metric_presure_comps}
\eeq
\ees
It is important to stress that the components in Eqs. (\ref{eq:boosted_metric_presure_comps1}, \ref{eq:boosted_metric_presure_comps}) are still those of flat spacetime, in a coordinate system that allowed its parameterization under the constant pressure $p$ and velocity $v_i$. It is, therefore, an exact solution to the vacuum Einstein's field equations. Nevertheless, this particular parameterization is most appropriate to perform the hydrodynamic expansion to the metric.

\subsection{The hydrodynamic expansion}
\label{sec:he1}

The constant velocity $v_i$ and pressure $p$ can be promoted to spacetime-dependent fields, using the hydrodynamic expansion to yield a near-equilibrium configuration. The procedure is similar to what is implemented in Ref. \cite{minwalla}, although a different kind of expansion will be performed.

By promoting $v_i$ and $p$ to arbitrarily $x^a$-dependent fields in Eqs. (\ref{eq:boosted_metric_presure_comps1}, \ref{eq:boosted_metric_presure_comps}),
\textcolor{black}{the resulting metric will be no longer a solution of the vacuum Einstein's field equations.}
The metric components carry an arbitrary spacetime dependence which does not, in general, comply with Einstein's field equations. 
On the other hand, if one promotes $v_i$ and $p$ to slowly-varying fields, with an amplitude small enough to be seen as perturbations around a background in which the vacuum Einstein's field equations are exactly solved, one can perturbatively solve the vacuum Einstein's field equations in the chosen perturbation parameter. To achieve this, the hydrodynamic expansion \bltx{of the fluid quantities is employed}, 
\bes
\beq
 v_i &\mapsto& v_i(x^a) = \bltx{\mathring{v_i}}(x^a), \label{eq:promotion1}
 \\
 p &\mapsto& p(x^a) = \frac{1}{\sqrt{r_c}} \left ( 1 + \frac{\bltx{\mathring{p}}(x^a)}{r_c} \right ),
 \label{eq:promotion}
\eeq\ees
\noindent where $\mathring{v_i}$ and $\mathring{p}$ are the $\epsilon$-parameterized quantities, expressing the hydrodynamic scaling of $v^i$ and $p$, as discussed in Sec. \ref{sec:hydro}.  
 Eqs. (\ref{eq:promotion1}, \ref{eq:promotion}) can be seen as a promotion of $v_i$ and $p$ to spacetime fields composed of small fluctuations, in the hydrodynamic expansion sense, around the equilibrium background $v_i = 0$, $p = r_c^{-1/2}$. According to Eq. \eqref{eq:by=pf_1}, it is the configuration reproducing the Rindler spacetime as an exact solution to the Einstein's field equations.

Therefore, since perturbations scaled under the hydrodynamic parameter $\epsilon$ are considered, one can perform the hydrodynamic expansion to the metric components (\ref{eq:boosted_metric_presure_comps1}, \ref{eq:boosted_metric_presure_comps}) after the mappings (\ref{eq:promotion1}, \ref{eq:promotion}). The resulting metric under this expansion perturbatively solves the vacuum Einstein's field equations  up to the desired order in $\epsilon$. This construction was extended to arbitrary order in $\epsilon$ \cite{fg3}, whereas in this work the expansion up to $\mathcal{O}(\epsilon^2)$ will be considered. 
The hydrodynamic expansion yields 
$
r_h = 2\bltx{\mathring{p}} + \mathcal{O}(\epsilon^4).
$ 
Since $\delta^{ij}\bltx{\mathring{v_i}}\bltx{\mathring{v_j}} \equiv \bltx{\mathring{v}^2} \sim \mathcal{O}(\epsilon^2)$, one obtains 
\begin{equation}
\begin{gathered}
\gamma^2 = \left (1-\frac{\bltx{\mathring{v}^2}}{r_c}\right )^{-1} = 1 + \frac{\bltx{\mathring{v}^2}}{r_c} + \mathcal{O}(\epsilon^4).
\end{gathered} 
\end{equation}


Now, the hydrodynamic expansion on the metric components will be implemented, for each component of Eqs. (\ref{eq:boosted_metric_presure_comps1}, \ref{eq:boosted_metric_presure_comps}), considering Eqs. (\ref{eq:promotion1}, \ref{eq:promotion}) and the expansion of the quantities. The resulting metric 
reads 
\begin{equation}
\begin{aligned}
g_{\mu \nu} \mathrm{d}x^\mu \mathrm{d}x^\nu = 
&-r \mathrm{d}t^2 + 2 \mathrm{d}t \mathrm{d}r +
\delta_{ij}\mathrm{d}x^i\mathrm{d}x^j 
-2 \left ( 1 - \frac{r}{r_c} \right ) \bltx{\mathring{v_i}} \mathrm{d}x^i \mathrm{d}t 
- \frac{2}{r_c} \bltx{\mathring{v_i}} \mathrm{d}x^i \mathrm{d}r +
\frac{\left ( \bltx{\mathring{v}}^2 + 2 \bltx{\mathring{p}} \right )}{r_c}\mathrm{d}t \mathrm{d}r 
\\
&+ \left ( 1 - \frac{r}{r_c}\right ) \left ( \bltx{\mathring{v}}^2 + 2 \bltx{\mathring{p}} \right ) \mathrm{d}t^2 
+ \left ( 1 - \frac{r}{r_c}\right ) \left ( \frac{1}{r_c} \bltx{\mathring{v_i}}\bltx{\mathring{v_j}} \mathrm{d}x^i \mathrm{d}x^j \right )+ \mathcal{O}(\epsilon^3) \ .
\label{eq:metric_expansion}
\end{aligned}
\end{equation}
The boundary condition of Eq. \eqref{eq:bc1_down} is met at $\Sigma_c$. 

The Brown--York stress tensor associated to the metric \eqref{eq:metric_expansion} can now be computed, first with the extrinsic curvature on $\Sigma_c$, 
\beq
 \!\!\!\!\!\!K_{ab}\mathrm{d}x^a \mathrm{d}x^b &\!=\!& -\frac{\left ( \bltx{\mathring{p}} \!+\! \bltx{\mathring{v}}^2 \!+\! r_c\right )}{2\sqrt{r_c}} \mathrm{d}t^2 \!+\! \frac{1}{\sqrt{r_c}}\bltx{\mathring{v_i}} \mathrm{d}x^i \mathrm{d}t+ \frac{1}{2\sqrt{r_c}} \left (\partial_{(i} \bltx{\mathring{v}}_{j)} - \frac{1}{r_c} \bltx{\mathring{v_i}} \bltx{\mathring{v_j}} \right ) \mathrm{d}x^i \mathrm{d}x^j + \mathcal{O}(\epsilon^3), 
\eeq
\noindent whose trace reads $K = \frac{1}{\sqrt{r_c}} \left (\frac{\bltx{\mathring{p}} + r_c}{2r_c} + \partial^i \bltx{\mathring{v_i}} \right ) + \mathcal{O}(\epsilon^4)$. Therefore, Eq. \eqref{eq:brown_york} yields 
\beq
T_{a b}^{\scalebox{.55}{BY}} \mathrm{d}x^a \mathrm{d}x^b &=&
\frac{1}{\sqrt{r_c}} \delta_{ij} \mathrm{d}x^i \mathrm{d}x^j 
- \frac{2}{\sqrt{r_c}} \bltx{\mathring{v_i}} \mathrm{d}x^i \mathrm{d}t + \frac{ \bltx{\mathring{v}^2} }{\sqrt{r_c}} \mathrm{d}t^2 
+ \frac{\left( \bltx{\mathring{v_i}} \bltx{\mathring{v_j}} + \bltx{\mathring{p}} \delta_{ij} - r_c\partial_{(i}\bltx{\mathring{v}}_{j)}\right)}{r_c^{3/2}} \mathrm{d}x^i \mathrm{d}x^j \n
&&
\qquad\qquad\qquad\qquad\qquad\qquad+ \frac{2}{\sqrt{r_c}} \partial^i \bltx{\mathring{v_i}} \gamma_{ab} \mathrm{d}x^a \mathrm{d}x^b +\mathcal{O}(\epsilon^3) \ .
\eeq

\subsection{The dual fluid}
\label{sec:df1}

Consider the conservation of the Brown--York stress tensor on $\Sigma_c$, 
\begin{equation}
 \nabla^a T_{ab} = 0.
 \label{eq:by_cons1}
\end{equation}
\noindent 
Notice that Eq. \eqref{eq:by_cons1} represents a constraint on $\Sigma_c$ which necessarily must be satisfied to ensure that the Einstein's field equations are solved perturbatively in $\epsilon$. In that sense, the conservation of the Brown--York stress tensor can be seen as an integrability condition of the Einstein's field equations \cite{fg3}. Besides, once the constraints are satisfied on $\Sigma_c$, it is possible to evolve the solution in the radial direction, provided the absence of singularities \cite{fg1, fg2}. Therefore, the Einstein's field equations can be reduced to the imposed Eq. \eqref{eq:by_cons1}. At $\mathcal{O}(\epsilon^2)$ order, $ \nabla^a T_{ab} = \gamma^{tt}\partial_t T_{tt} + \gamma^{ij} \partial_j T_{it} + \mathcal{O}(\epsilon^4) = 0$, reducing to
\begin{equation}
\begin{gathered}
\partial^i\bltx{\mathring{v_i}} \sim \mathcal{O}(\epsilon^4),
\label{eq:incomp_1}
\end{gathered}
\end{equation}
\noindent which is satisfied by an incompressible fluid, up to $\mathcal{O}(\epsilon^4)$. On the other hand, at $\mathcal{O}(\epsilon^3)$ one has $\nabla^a T_{aj} = \gamma^{tt}\partial_t T_{tj} + \gamma^{ik} \partial_k T_{ij} + \mathcal{O}(\epsilon^4) \sim 0 $, which yields
\begin{equation}
\begin{gathered}
\partial_t \bltx{\mathring{v_j}} - \eta \partial^2 \bltx{\mathring{v_j}} + \partial_j \bltx{\mathring{p}} + \bltx{\mathring{v}^i} \partial_i \bltx{\mathring{v_j}} \sim \mathcal{O}(\epsilon^4),
\label{eq:ns_1}
\end{gathered}
\end{equation}
\noindent where the kinematic viscosity $\eta = r_c$ has been identified, also using Eq. \eqref{eq:incomp_1} 
to obtain the incompressible Navier--Stokes equations for a fluid of velocity field $ \bltx{\mathring{v_i}}$ and pressure field $ \bltx{\mathring{p}}$.

This is precisely the sense in which the vacuum Einstein's field equations are reduced to the incompressible Navier--Stokes equations, within the setup and boundary conditions considered. One can then assert that there is an incompressible fluid on $\Sigma_c$, whose velocity and pressure fields parameterize the bulk spacetime, according to Eq. \eqref{eq:metric_expansion}. It is worth mentioning that Ref. \cite{fg2} performs a near-horizon expansion of the Rindler spacetime initially considered, by taking the acceleration of $\Sigma_c$ to infinity, as this results in pushing it to its future horizon. It is then shown that the near-horizon expansion is formally equivalent to the hydrodynamic expansion. Therefore, in the near-horizon limit, one can see $\Sigma_c$ located at the horizon, thus establishing the precise sense in which the horizon --- and not only an arbitrary manifold at constant radius --- can be seen as an incompressible fluid. 

\ptx{The dual fluid obtained through this procedure presents remarkable properties, namely, a vanishing energy density and non-zero pressure in equilibrium, which is indeed the case up to $\mathcal{O}(\epsilon^3)$. Sec. \ref{sec:fixed} discusses and address this property. However, for the fluid description to be complete, an equation of state must be chosen to solve the equations of motion. Ref. \cite{fg3} extensive analyzes the dual fluid, discussing how the hydrodynamic description must change to accommodate a vanishing equilibrium energy density.} 
\ptx{First, a Hamiltonian constraint is then introduced on $\Sigma_c$ and imposed to the Brown-York stress tensor of the dual fluid,
\begin{equation}
\label{Ham_constraint}
T^2 = D T_{ab}T^{ab} \ . 
\end{equation}}
\!\ptx{Indeed, the application of this constraint to equilibrium configurations of the fluid is responsible for fixing: a) $\rho=0$, which is the case here scrutinized; or b) $\rho=\left(-2D/(D-1)\right) p$. The latter has been not recovered yet, by the current present procedure. Instead, the more unexpected result $\rho=0$ is what we have at hand. As will be seen below, these results emerge from an equation that may be seen as an equation of state.}

\ptx{If one considers dissipative terms, the stress tensor of a perfect fluid, expressed in Eq. \eqref{eq:perfect_fluid}, is modified to
\begin{equation}
\label{hydroT}
 T_{ab} = \rho u_a u_b + p \left ( \gamma_{ab}+u_a u_b \right) + \Pi^\perp_{ab}, \qquad\qquad u^a\Pi^\perp_{ab}=0 \ , 
\end{equation}
\noindent where $\Pi^\perp_{ab}$ encodes dissipative corrections, often expanded in the gradient of the fluid velocity. The Landau gauge condition was also considered, which allows the fluid momentum density to vanish in the local rest frame \cite{landau_fluids}.
Now, given Eq. \eqref{eq:by=pf_1}, one can replace Eq. \eqref{hydroT} into the Hamiltonian constrain in Eq. \eqref{Ham_constraint}, yielding
\begin{equation}
\label{rho_result}
\rho \big( (D-1)\rho+2D p+2\Pi^{\perp}\big) + D\Pi^\perp_{ab}\Pi^{\perp ab}-(\Pi^\perp)^2=0,
\end{equation}
\noindent where $\Pi^\perp = h^{ab}\Pi^\perp_{ab}$, and $h_{ab} = \gamma_{ab} u_a u_b$.
This relation fully determines the energy density $\rho$ in terms of $p$ and $\Pi^\perp_{ab}$. Therefore, Eq. \eqref{rho_result} plays the role of the equation of state for the fluid. Indeed, in the absence of dissipative terms, either $\rho=0$ or $\rho=\left(-2D/(D-1)\right) p$ are both recovered. Finally, Ref. \cite{Bhattacharyya:2008kq} addressed that Navier--Stokes equations can be described by the scaling limit of relativistic equations of fluid dynamics, whose dual fluid dynamical theory is conformal. Also, the conformal group descends to symmetries of
the Navier--Stokes equations. Nevertheless, a technical point must be mentioned. The non-relativistic conformal symmetry group of the Navier--Stokes equations comprises spatial conformal generators $K_i$, however a generator that descends from the temporal conformal generator $K_0$ is not included, thus generating a group that is indeed distinct of the Schr\"odinger group \cite{Son:2008ye}.} 

\ptx{Therefore, the equation of state \eqref{rho_result} does not necessarily indicate a conformal fluid, which by no means is a novel result in this work since it was previously presented in the literature \cite{fg3}. Further, as aforementioned, Ref. \cite{Bhattacharyya:2008kq} stated that the procedure implemented here is indeed valid for arbitrary equations of motions so that a conformal invariance of the parent hydrodynamical system is only a particular choice, instead of a necessary condition.}

\section{A varying induced metric}
\label{sec:varying}

\bltx{Consider now a metric in $(D+2) \geq 3$ dimensions with a non-extremal horizon endowing the spacetime, covered by the chart $\left \{\tau, \uprho, x^i\right \}$}, where $x^i$ are spatial coordinates transverse to the horizon. Within the assumption that the horizon is free of singularities, a Taylor expansion is allowed in the near-horizon region, with the radius $\uprho$ as the expansion parameter, such that the horizon is placed at $\uprho = 0$. The following near-horizon boundary conditions are adopted \cite{daniel1, daniel2},
\beq
g_{\tau \tau} = - \kappa^2 \uprho^2, \qquad g_{\uprho \uprho} = 1 + \mathcal{O}(\uprho^2),\qquad
g_{\uprho i} = \mathcal{O}(\uprho),\qquad g_{ij} = \Omega_{ij} + \mathcal{O}(\uprho^2),
\label{eq:bc_nh}
\eeq
\noindent where $\kappa$ is the Rindler acceleration, taken to be greater than zero to ensure non-extremality, whereas $\Omega_{ij} = \Omega_{ij}(t, x^i)$ is the metric transverse to the horizon, such that $\det \Omega_{ij} \neq 0$, avoiding singularities. \bltx{Ref. \cite{daniel3} demonstrated that the symmetries of $\Omega_{ij}$ are generated by an algebra that is the semi-direct sum of diffeomorphisms at the $(D - 2)$-dimensional
spacelike section of the horizon and a generalization of supertranslations. This algebra emulates the Bondi--van der Burg--Metzner--Sachs (BMS) algebra for this Rindler-like space. }An important special case is when the boundary metric $\Omega_{ij}$ does not depend explicitly on time. It corresponds to a constant surface gravity, which is the case of interest here. Therefore, we will consider a time-independent boundary metric.

These near-horizon boundary conditions lead to an infinite set of near horizon symmetries and associated soft hair excitations \cite{minwalla,daniel3,sh,Donnay:2015abr,Donnay:2016ejv,phdthesis}, featuring in any non-extremal horizon, which holds for any spacetime dimension greater than two. The main goal of this section is then to investigate the relationship between fluids and the soft hairy horizon generated by this choice of boundary conditions within the same general setup described in Sec. \ref{sec:setup}.
It is important to notice that the boundary conditions (\ref{eq:bc_nh}) were specifically introduced to constrain the $(D+2)$-dimensional metric in the near-horizon expansion. However, as discussed in Sec. \ref{sec:df1}, the near-horizon and hydrodynamic expansions are formally identical. For this reason, one can emulate the procedure adopted in the last section for the fixed induced metric, performing the hydrodynamic expansion also for these new boundary conditions. In practice, we will perform a hydrodynamic expansion considering the boundary condition $g_{ij} = \Omega_{ij}$ to hold on a generic manifold $\Sigma_c$, which in the near-horizon limit will be identified to the horizon.

For computational purposes, we choose as a boundary condition for the $(D+2)$-dimensional spacetime the following induced metric on $\Sigma_c$, 
\begin{equation}
\gamma_{a b} \mathrm{d}x^a \mathrm{d}x^b = -r_c \mathrm{d} t^2 + \Omega_{ij}\mathrm{d}x^i \mathrm{d}x^j, 
\label{eq:bc2_down}
\end{equation}
\noindent where $\Omega_{ij} = \Omega_{ij}(x^k)$ is the $D$-dimensional boundary metric of $\Sigma_c$. Although the physical interpretation here in principle differs from the one in Ref. \cite{daniel3}, the near-horizon/hydrodynamic equivalence  formally ensures the same interpretation.

The procedure in the previous section must be then emulated with this important change in the boundary conditions, which leads to interesting generalizations. Considering a $(D+2)$-dimensional spacetime covered by ingoing Rindler coordinates $\{r, t, x^i\}$, it cannot be regarded as the Rindler spacetime itself, since now $\Omega_{ij}$ determines the spacelike part of the metric, 
\begin{equation}
 g_{\mu \nu} \mathrm{d}x^\mu \mathrm{d}x^\nu = -r \mathrm{d}t^2 + 2 \mathrm{d}t \mathrm{d}r + \Omega_{ij}\mathrm{d}x^i \mathrm{d}x^j.
 \label{eq:rindler_st_new}
\end{equation} 
\bltx{However, we impose that the metric \eqref{eq:rindler_st_new} exactly solves the vacuum Einstein's field equations. Naturally, this imposition constraints the functional dependence of $\Omega_{ij}$. Nonetheless, for the relevant metrics which lead to soft hair excitations, this is indeed the case. In Ref. \cite{daniel3}, the Kerr--Taub--NUT metric in Boyer--Lindquist coordinates is studied as a concrete example, for which case an explicit functional dependence of $\Omega_{ij}$ is presented. See Appendix \ref{ap:kerr}.}

\textcolor{black}{By following an analogous procedure to the one in Sec. \ref{sec:fixed}, one straightforwardly finds that}
%
\begin{equation}
T_{a b}^{\scalebox{.55}{BY}} = T_{ab}^{\scalebox{.55}{PF}}.
\label{eq:by=pf_quase_1}
\end{equation}
Notice that one has the same background as before, in which the Brown--York stress tensor is precisely equal to that of a perfect fluid with vanishing energy density, just like in Eq. \eqref{eq:by=pf_1}. That is of major importance, as it allows the hydrodynamical expansion, which will be implemented in what follows.

\subsection{Boosting the metric}

A family of $(D+2)$-dimensional metrics is now considered, parameterized by the constant velocity $v_i$, which is achieved by the diffeomorphisms presented in Eq. \eqref{eq:rindler_st_new}, collectively referred as boosts, although they consist of rescales and shifts as well. Now, since these diffeomorphisms do not involve any derivatives, it is clear that the resulting boosted metric components have the same form as in Eqs. (\ref{eq:boosted_metric_comps1}, \ref{eq:boosted_metric_comps}), but with the replacement $\delta_{ij} \mapsto \Omega_{ij}(x^k)$,
\bes
\beq
g_{tt} &=& \gamma^2 \left ( v^2 - \alpha^2 \left ( r - r_h \right ) \right ), \qquad \quad
g_{tr} = \gamma \alpha, \qquad\quad 
g_{ti} = \frac{\gamma^2 \alpha^2}{r_c} \left ( r - r_c \right ) v_i, \label{eq:boosted_metric_comps_1}\\ 
g_{rr} &=& 0, \qquad\quad g_{ri} = -\frac{\gamma \alpha}{r_c}v_i, \qquad\quad  
g_{ij} = \Omega_{ij} - \frac{\gamma^2 \alpha^2}{r_c^2} \left (r -r_c \right ) v_i v_j.
\label{eq:boosted_metric_comps_2}
\eeq
\ees
Therefore, \textcolor{black}{the} extrinsic curvature on $\Sigma_c$ is given by \begin{equation}
\begin{aligned}
\!\!\!\!\!K_{ab}\mathrm{d}x^a \mathrm{d}x^b \!=\! \frac{\gamma}{2\sqrt{r_c}} v^k \partial_k \Omega_{ij} \mathrm{d}x^i \mathrm{d}x^j \!+\! \gamma^2 \alpha \frac{\sqrt{r_c}}{2} \left ( \frac{2}{r_c}v_i \mathrm{d}x^i \mathrm{d}t \right )
\!-\! \gamma^2 \alpha \frac{\sqrt{r_c}}{2} \left ( \mathrm{d}t^2 \!+\! \frac{1}{r_c^2} v_i v_j \mathrm{d}x^i \mathrm{d}x^j \right ),
\end{aligned}
\end{equation}
\noindent whose trace is $K = \frac{1}{2\sqrt{r_c}} \left ( \alpha + \gamma \Omega^{ij} v^k \partial_k \Omega_{ij} \right )$. From this, it is possible to compute the Brown--York stress tensor, according to Eq. \eqref{eq:brown_york},
\beq
T_{a b}^{\scalebox{.55}{BY}} \mathrm{d}x^a \mathrm{d}x^b &=& \frac{\alpha}{\sqrt{r_c}} \left [ \gamma^2 v^2 \mathrm{d}t^2 -2\gamma^2 v_i \mathrm{d}x^i \mathrm{d}t\right ] + \frac{\alpha}{\sqrt{r_c}} \left ( \Omega_{ij} + \frac{\gamma^2}{r_c}v_i v_j \right ) \mathrm{d}x^i\mathrm{d}x^j
\non\\
&&+ \frac{\gamma}{\sqrt{r_c}} \left ( \Omega^{kl} v^m \partial_m \Omega_{kl} \Omega_{ij} - v^k \partial_k \Omega_{ij} \right ) \mathrm{d}x^i \mathrm{d}x^j- \gamma \sqrt{r_c} \Omega^{ij} v^k \partial_k \Omega_{ij} \mathrm{d}t^2.
\eeq
Now considering a perfect fluid with \bes
\beq\rho &=& -\gamma \Omega^{ij}v^k \partial_k \Omega_{ij} r_c^{-1/2},\\
p &=& \left( \alpha + \gamma \Omega^{ij}v^k \partial_k \Omega_{ij} \right ) r_c^{-1/2},\\
u_a &=& \gamma \left( -r_c^{1/2} , r_c^{-1/2} v_i\right),\eeq\ees its stress tensor is, according to Eq. \eqref{eq:perfect_fluid}, given by
\beq
T_{a b}^{\scalebox{.55}{PF}} \mathrm{d}x^a \mathrm{d}x^b &=& \frac{\alpha}{\sqrt{r_c}} \left [ \gamma^2 v^2 \mathrm{d}t^2 -2\gamma^2 v_i \mathrm{d}x^i \mathrm{d}t\right ] - \gamma \sqrt{r_c} \Omega^{ij} v^k \partial_k \Omega_{ij} \mathrm{d}t^2 + \frac{\alpha}{\sqrt{r_c}} \left ( \Omega_{ij} + \frac{\gamma^2}{r_c}v_i v_j \right ) \mathrm{d}x^i\mathrm{d}x^j\nonumber
\\
&&\qquad\qquad\qquad\qquad\qquad\qquad\qquad\qquad+ \frac{\gamma}{\sqrt{r_c}} \left ( \Omega^{kl} v^m \partial_m \Omega_{kl} \Omega_{ij} \right ) \mathrm{d}x^i \mathrm{d}x^j.
\eeq
Therefore, the Brown--York stress tensor has the form of a perfect fluid on $\Sigma_c$ with a correction,
\begin{equation}
T_{a b}^{\scalebox{.55}{BY}} = T_{ab}^{\scalebox{.55}{PF}} + \tilde{T}_{ab},
\label{eq:by=pf_quase_2}
\end{equation}
\noindent where 
\begin{equation}
 \tilde{T}_{ab}\mathrm{d}x^a \mathrm{d}x^b = -\frac{\gamma}{\sqrt{r_c}}v^k \partial_k \Omega_{ij} \mathrm{d}x^i \mathrm{d}x^j.
\end{equation}
Hence the new boundary conditions yield a fluid with a non-zero energy density $\rho$ \bltx{--- although this will be corrected after the imposition of the relevant boundary condition, below}. The changes in the other parameters of the fluid, compared to the results for the non-boosted metric, once again are a direct consequence of the diffeomorphisms utilized.

Motivated by the hydrodynamic expansion, the fluid pressure can be introduced as a parameter of the metric components, by substituting $\alpha = \sqrt{r_c} \left ( p + \rho \right) $ in Eqs. (\ref{eq:boosted_metric_comps_1}, \ref{eq:boosted_metric_comps_2}):
\bes
\beq
\!\!\!\!\!\!\!\!\!\!\!\!g_{tt} &\!=\!& \gamma^2 \left ( v^2 \!-\! r_c\left(p + \rho \right)^2\left ( r \!-\! r_h \right ) \right ),\qquad\qquad
g_{tr} \!=\! \gamma \sqrt{r_c} \left(p \!+\! \rho \right),\qquad \label{eq:boosted_metric_presure_comps_1}\\ 
\!\!\!\!\!\!\!\!\!\!\!\!g_{rr} &=& 0,\qquad\qquad\qquad\qquad\qquad\qquad\qquad
g_{ti} \!=\! \gamma^2 \left(p \!+\! \rho \right)^2 \left ( r \!-\! r_c \right ) v_i,\label{eq:boosted_metric_presure_comps_2}\\ 
\!\!\!\!\!\!\!\!\!\!\!\!g_{ri} &=& -\frac{\gamma}{\sqrt{r_c}}\left(p + \rho \right)v_i,\qquad\qquad 
g_{ij} = \Omega_{ij} - \frac{\gamma^2}{r_c} \left(p + \rho \right)^2 \left (r -r_c \right ) v_i v_j.
\label{eq:boosted_metric_presure_comps_3}
\eeq\ees
Now we are ready to promote $v_i$ and $p$ to spacetime fields under the hydrodynamic scaling.

\subsection{The hydrodynamic expansion}
\label{sec:he2}

The constant velocity $v_i$ and pressure $p$ can be now promoted to slowly varying $x^a$-dependent fields with an amplitude small enough as to be seen as perturbations around a background wherein the vacuum Einstein's field equations are exactly solved,
\begin{equation}
\begin{aligned}
 v_i &\mapsto v_i(x^a) = \bltx{\mathring{v_i}}(x^a),  
 \qquad\quad p \mapsto p(x^a) = \frac{1}{\sqrt{r_c}} \left ( 1 + \frac{\bltx{\mathring{p}}(x^a)}{r_c} \right ).
\end{aligned}
\end{equation}
Here again, the spacetime fields $v_i$ and $p$ are composed of small (in the hydrodynamic expansion sense) fluctuations around the equilibrium background $v_i = 0$, $p = r_c^{-1/2}$, which, according to Eq. \eqref{eq:by=pf_quase_1}, is the configuration reproducing Eq. \eqref{eq:rindler_st_new} as an exact solution to the Einstein's field equations. 

The hydrodynamic expansion yields 
\beq 
r_h &=& 2\bltx{\mathring{p}} + \mathcal{O}(\epsilon^4),\\
\gamma^2 &=& \left (1-\frac{\bltx{\mathring{v}^2}}{r_c}\right )^{-1} = 1 + \frac{\bltx{\mathring{v}^2}}{r_c} + \mathcal{O}(\epsilon^4),
\eeq
\noindent where $\Omega^{ij}\bltx{\mathring{v_i}}\bltx{\mathring{v_j}} \equiv \bltx{\mathring{v}^2} \sim \mathcal{O}(\epsilon^2)$. Consequently, 
\begin{equation}
\begin{aligned}
\rho &= -\frac{1}{\sqrt{r_c}} \Omega^{ij} \bltx{\mathring{v_k}} \partial_k \Omega_{ij} + \mathcal{O}(\epsilon^4) \ ,
\end{aligned}
\end{equation}
\noindent so that
\begin{equation}
(p + \rho)^2 = \frac{1}{r_c} \left ( 1 + \frac{2 \bltx{\mathring{p}}}{r_c} - 2\Omega^{ij}\bltx{\mathring{v}^k} \partial_k\Omega_{ij} \right ) + \mathcal{O}(\epsilon^4).
\label{eq:p+rho^2}
\end{equation}
The hydrodynamic expansion on the metric components of Eqs. (\ref{eq:boosted_metric_presure_comps_1} - \ref{eq:boosted_metric_presure_comps_3}) yields the metric\beq
g_{\mu \nu} \mathrm{d}x^\mu \mathrm{d}x^\nu \!&\!=\! &\!-r \mathrm{d}t^2 \!+\! 2 \mathrm{d}t \mathrm{d}r \!+\! \Omega_{ij}\mathrm{d}x^i\mathrm{d}x^j \!-\!2 \left ( 1 \!-\! \frac{r}{r_c} \right ) \bltx{\mathring{v_i}} \mathrm{d}x^i \mathrm{d}t - \frac{2}{r_c}\bltx{\mathring{v_i}} \mathrm{d}x^i \mathrm{d}r + \left ( 1 \!-\! \frac{r}{r_c}\right ) \left (\bltx{\mathring{v}^2} + 2\textcolor{black}{\bltx{\mathring{p}}} \right ) \mathrm{d}t^2\n
&\!\!\!+\!\!&\!\!\! \left ( 1 \!-\! \frac{r}{r_c}\right ) \!\frac{\bltx{\mathring{v_i}} \bltx{\mathring{v_j}}}{r_c} \mathrm{d}x^i \mathrm{d}x^j\!+\! \frac{\left (\bltx{\mathring{v}^2} \!+\! 2\textcolor{black}{\bltx{\mathring{p}}} \right )}{r_c}\mathrm{d}t \mathrm{d}r
 \!+\! 2r \Omega^{ij}\bltx{\mathring{v}^k} \partial_k\Omega_{ij} \mathrm{d}t^2 \!-\! 2\Omega^{ij}\bltx{\mathring{v}^k} \partial_k\Omega_{ij} \mathrm{d}t \mathrm{d}r \!+\! \mathcal{O}(\epsilon^3).
\label{eq:metric_expansion_2}
\eeq
\noindent From this, the induced metric on $\Sigma_c$ reads
\begin{equation}
\begin{aligned}
\gamma_{a b} \mathrm{d}x^a \mathrm{d}x^b &= -r_c \left( 1 - 2\Omega^{ij}\bltx{\mathring{v}^k} \partial_k\Omega_{ij} \right )\mathrm{d} t^2 
+ \Omega_{ij}\mathrm{d}x^i \mathrm{d}x^j + \mathcal{O}(\epsilon^3).
\label{eq:bc2_down_gen}
\end{aligned}
\end{equation}
Therefore, one must impose the boundary condition \eqref{eq:bc2_down}, meaning that $g_{tt}$ can be regarded as a constant up to $\mathcal{O}(\epsilon^3)$, \begin{equation}
\Omega^{ij}\bltx{\mathring{v}^k} \partial_k\Omega_{ij} \sim\mathcal{O}(\epsilon^3),
\label{eq:salvador}
\end{equation}
\noindent so that the induced metric is effectively given by \begin{equation}
\gamma_{a b} \mathrm{d}x^a \mathrm{d}x^b = -r_c \mathrm{d} t^2 + \Omega_{ij}\mathrm{d}x^i \mathrm{d}x^j + \mathcal{O}(\epsilon^3), 
\end{equation}
\noindent exactly as imposed by Eq. \eqref{eq:bc2_down}, up to $\mathcal{O}(\epsilon^3)$.

Notice that the constraint of Eq. \eqref{eq:salvador} appeared after the imposition of the chosen boundary conditions, being, therefore, of major importance for consistency. Also, it could be relaxed to $\Omega^{ij}\bltx{\mathring{v}^k} \partial_k\Omega_{ij} = \text{constant}$, and $\kappa$ would be still a constant. However, to satisfy the specific boundary condition in Eq. \eqref{eq:bc2_down}, and to fix the unjustified energy density acquired by the perfect fluid after the boost, the constant must be set to zero. Notice that Eq. \eqref{eq:salvador} may be seen as a tensor density equation, which is straightforwardly promoted to a covariant expression, after the multiplication by the weight at both sides. \bltx{At first, this condition seems to imply a constraint on the allowed directions of the velocity field $\bltx{\mathring{v}^k}$, however, further investigation of its physical content is to be done in future works.}

Applying the constraint of Eq. \eqref{eq:salvador}, some of the results already calculated can be simplified. Eq. \eqref{eq:p+rho^2} reduces to
\begin{equation}
(p + \rho)^2 = \frac{1}{r_c} \left ( 1 + \frac{2\textcolor{black}{\bltx{\mathring{p}}}}{r_c} \right ) + \mathcal{O}(\epsilon^4).
\end{equation}
 It may be useful to consider a slightly different way to calculate the extrinsic curvature of that presented in Eq. \eqref{eq:ext_curv}. Since the Lie derivative is connection independent, one may simply substitute the partial derivatives by covariant derivatives, yielding precisely the same result. Notice, however, that due to the metric compatibility condition, $\nabla_\mu \gamma_{\nu \sigma} = 0$, the first term in the extrinsic curvature vanishes, and we are left with $K_{\mu \nu} = \frac{1}{2} \left (\gamma_{\sigma \nu} \nabla_\mu n^\sigma + \gamma_{\mu \sigma} \nabla_\nu n^\sigma \right )$. 
 

Calculating the connection coefficients associated with $\gamma_{\mu \nu}$ the extrinsic curvature on $\Sigma_c$ reads
\beq
\!\!\!\!\!\!\!\!\!\!\!\!\!\!\!\!K_{ab}\mathrm{d}x^a \mathrm{d}x^b &=& -\frac{\left ( \textcolor{black}{\bltx{\mathring{p}}} + \bltx{\mathring{v}^2} + r_c\right )}{2\sqrt{r_c}} \mathrm{d}t^2 + \frac{1}{\sqrt{r_c}} \bltx{\mathring{v}_i} \mathrm{d}x^i \mathrm{d}t+\frac{1}{2\sqrt{r_c}} \left ( \nabla_{(i} \bltx{\mathring{v}_{j)}} - \frac{1}{r_c} \bltx{\mathring{v}_i} \bltx{\mathring{v}_j} \right ) \mathrm{d}x^i \mathrm{d}x^j 
+ \mathcal{O}(\epsilon^3),
\eeq
\noindent whose trace is $K = \frac{1}{2\sqrt{r_c}} \left (\frac{\textcolor{black}{\bltx{\mathring{p}}} + r_c}{r_c} + 2\nabla^i \bltx{\mathring{v}_i} \right ) + \mathcal{O}(\epsilon^4)$. Hence, the Brown--York stress tensor on $\Sigma_c$ is given by 
\beq
\!\!\!\!\!\!\!\!\!\!\!T_{a b}^{\scalebox{.55}{BY}} \mathrm{d}x^a \mathrm{d}x^b &=& \frac{1}{\sqrt{r_c}} \Omega_{ij} \mathrm{d}x^i \mathrm{d}x^j 
- \frac{2}{\sqrt{r_c}} \bltx{\mathring{v}_i} \mathrm{d}x^i \mathrm{d}t + \frac{ \bltx{\mathring{v}^2}}{\sqrt{r_c}} \mathrm{d}t^2 + \frac{1}{r_c^{3/2}} \left ( \bltx{\mathring{v}_i} \bltx{\mathring{v}_j} + \textcolor{black}{\bltx{\mathring{p}}} \Omega_{ij} \right ) \mathrm{d}x^i \mathrm{d}x^j \n&& 
- \frac{1}{\sqrt{r_c}} \left (\nabla_i \bltx{\mathring{v}_j} + \nabla_j \bltx{\mathring{v}_i} \right ) \mathrm{d}x^i \mathrm{d}x^j + \frac{2}{\sqrt{r_c}} \nabla^k \bltx{\mathring{v}_k} \Omega_{ij}\mathrm{d}x^i \mathrm{d}x^j - 2\sqrt{r_c} \nabla^i \bltx{\mathring{v}_i} \mathrm{d}t^2 + \mathcal{O}(\epsilon^3).
\eeq

\subsection{The dual fluid}
\label{sec:df2}

\bltx{To obtain the dual fluid, consider again the conservation of the Brown--York stress tensor on $\Sigma_c$,
\begin{equation}
 \nabla^a T_{ab} = 0.
 \label{eq:by_cons2}
\end{equation}
As before, imposing the conservation of the Brown--York stress tensor yields the Navier--Stokes equations. For functional dependencies in $\Omega_{ij}$ that satisfy Einstein's field equations, this condition always holds,} and may be seen as an integrability condition of Einstein's field equations, also representing a constraint on $\Sigma_c$ which necessarily must be satisfied 
\textcolor{black}{to guarantee that Einstein's field equations are solved perturbatively in $\epsilon$.} Once the constraint of Eq. \eqref{eq:by_cons2} is satisfied on $\Sigma_c$, it is possible to evolve the solution in the radial direction, which precisely establishes the sense in which we can reduce the Einstein's field equations to Eq. \eqref{eq:by_cons2}. Therefore, at $\mathcal{O}(\epsilon^2)$, the relation $\nabla^a T_{at} = \gamma^{tt}\nabla_t T_{tt} + \gamma^{ij} \nabla_j T_{it} = 0$ holds, which yields
\begin{equation}
\begin{gathered}
\nabla^i \bltx{\mathring{v}_i} \sim \mathcal{O}(\epsilon^4).
\label{eq:incomp_2}
\end{gathered}
\end{equation}
Eq. \eqref{eq:incomp_2} states the incompressibility of the fluid up to $\mathcal{O}(\epsilon^4)$ in a covariant manner, thus being valid in the more general background we are considering. At $\mathcal{O}(\epsilon^3)$, one has $\nabla^a T_{aj} = \gamma^{tt}\nabla_t T_{tj} + \gamma^{ik} \nabla_k T_{ij} + \mathcal{O}(\epsilon^4) = 0$, which yields
\begin{equation}
\begin{gathered}
\nabla_t \bltx{\mathring{v}_j} - r_c \nabla^2 \bltx{\mathring{v}_j} + \nabla_j \textcolor{black}{\bltx{\mathring{p}}} + \bltx{\mathring{v}^i} \nabla_i \bltx{\mathring{v}_j}-r_c \nabla^i \nabla_j \bltx{\mathring{v}_i} \sim \mathcal{O}(\epsilon^4).
\end{gathered}
\end{equation}
The commutator of the covariant derivatives is such that
\begin{equation}
\begin{gathered}
\nabla^i \nabla_j \bltx{\mathring{v}_i} = R^{\,i}_{j} \bltx{\mathring{v}_i} + \nabla_j \overbrace{\nabla^i \bltx{\mathring{v}_i}}^{\mathcal{O}(\epsilon^4)} = R^{\,i}_{j} \bltx{\mathring{v}_i} + \mathcal{O}(\epsilon^4) \ ,
\end{gathered}
\end{equation}
\noindent where $R^{\,i}_{j}$ denote the Ricci tensor components. Therefore, it implies that 
\begin{equation}
\begin{gathered}
\nabla_t \bltx{\mathring{v}_j} - \eta \left (\nabla^2 \bltx{\mathring{v}_j} + R^{\,i}_{j} \bltx{\mathring{v}_i} \right ) + \nabla_j \textcolor{black}{\bltx{\mathring{p}}} + \bltx{\mathring{v}^i} \nabla_i \bltx{\mathring{v}_j} \sim \mathcal{O}(\epsilon^4),
\label{eq:ns_2}
\end{gathered}
\end{equation}
\noindent where $\eta = r_c$ can be identified to the kinematic viscosity of the fluid, whose velocity and pressure fields satisfy Eq. \eqref{eq:ns_2}. This is a manifestly covariant generalization of the incompressible Navier--Stokes equation, up to $\mathcal{O}(\epsilon^4)$, thus generalizing the description of the fluid located in the manifold featuring the imposed boundary conditions. 
\textcolor{black}{One can think of $R^{\,i}_{j} \bltx{\mathring{v}_i}$ in Eq. \eqref{eq:ns_2} as an effective term in the Navier--Stokes equation that corrects the Laplacian in the diffusion term, $\eta \left (\nabla^2 \bltx{\mathring{v}_j} + R^{\,i}_{j} \bltx{\mathring{v}_i} \right)$. This term can be also interpreted as the difference between the velocity at a point and the mean velocity in a small surrounding volume in the manifold featuring the imposed boundary conditions, now altered by a Ricci tensor contribution, given the non-flat background with soft hair excitations. This also implies that there is an effectively increased viscosity by diffusion, driven by $\Omega_{ij}$. It is also worth emphasizing that the convection term, 
$ \bltx{\mathring{v}^i} \nabla_i \bltx{\mathring{v}_j}$ remains unaltered and there is no additional fluid flow driven by eventual density differences, which complies with our hypotheses. Besides, the term proportional to the Ricci tensor is a direct consequence of the non-commutativity of the covariant derivatives, in a gravitational background with soft hair excitations.}
Therefore, there is a precise sense in which the vacuum Einstein's field equations are reduced to the generalized incompressible Navier--Stokes equations describing a fluid on $\Sigma_c$, which --- given the equivalence between the hydrodynamic and near-horizon expansions --- is identified with a soft hairy horizon.

\section{Concluding remarks}
\label{sec:conc}

In this work we have shown how to generalize the \textcolor{black}{particular realization of the fluid/gravity correspondence} in Ref. \cite{fg2} by considering a new set of horizon boundary conditions, which are associated with soft hair excitations. This was possible by employing the near-horizon hydrodynamic expansion equivalence of \cite{fg2}, which allowed us to employ these boundary conditions to the hydrodynamic expansion of the metric. A similar construction leads to the reduction of vacuum Einstein's field equations to generalized, manifestly covariant Navier--Stokes equations \eqref{eq:ns_2} on $\Sigma_c$. It generalizes the description of the fluid located in the manifold featuring the imposed boundary conditions. As a consequence, and given the fact that one can realize $\Sigma_c$ as being located at the horizon, it was possible to establish the precise sense in which the soft hairy horizon is a generalized incompressible fluid.  
\bltx{As asymptotic symmetries close to non-extremal black hole horizons are generated by an extension of supertranslations \cite{Donnay:2015abr}, one might also speculate whether Virasoro and Abelian currents play some role regarding generalizations of the results in this manuscript \cite{Donnay:2016ejv,Donnay:2020guq}.} 

As a perspective, fermionic modes can be considered in a supergravity setup, where 
the Brown--York energy-momentum tensor can be then evaluated, with fermionic corrections to the perfect fluid on the boundary \cite{Meert:2018qzk}. 
Ref. \cite{Gentile:2012jm} suggests fermionic wigs consisting of anticommuting counterparts of black hole hairs, considering AdS/CFT between the supergravity and
its dual theory yields supersymmetric extensions of Navier--Stokes equations. Therefore, fermionic corrections to Navier--Stokes can be derived when supersymmetric extensions of AdS--Schwarzschild black branes implement fermionic wigs. 
We intend to extend and emulate the metric solution and the Brown--York energy-momentum tensor in Refs. \cite{Gentile:2013gea,Gentile:2012jm} on the boundary to introduce the concept of a soft wig.

\subsection*{Acknowledgements}
\label{sec:thanks}

AJFM is grateful to FAPESP (Grants 2017/13046-0 and 2018/00570-5) and CAPES - Brazil for financial support.
RdR expresses deep gratitude to FAPESP (Grants No. 2017/18897-8 and No. 2021/01089-1) and the National Council for Scientific and Technological Development -- CNPq (Grants No. 303390/2019-0 and No. 406134/2018-9), for partial financial support.

\appendix
\section{A specific form of the boundary metric transverse to the horizon}
\label{ap:kerr}

As an example, one can consider the Kerr--Taub--NUT metric in Boyer--Lindquist coordinates 
\begin{align}
\mathrm{d} s^2 = -\frac{\Delta}{\Upgamma}
\big(\mathrm{d} {t}-(a \sin^2\theta -2n \cos\theta) \mathrm{d} \phi \big)^2 +\frac{\Upgamma}{\Delta}\mathrm{d} r^2
+\Upgamma \mathrm{d} \theta^2 +\frac{\sin^2\theta}{\Upgamma}\big(a \mathrm{d} {t}-(r^2+a^2+n^2)\mathrm{d} \phi \big)^2 
\nonumber
\end{align}
with
$
\Delta = r^2-2Mr+\kerr^2-n^2$ and $\Upgamma = r^2+(n+\kerr\cos\theta)^2$ and $n$ denotes the NUT charge. 
This metric has Killing horizons located at the zeros of $\Delta$, given by
$
r_{\pm}=M\pm\sqrt{M^2-\kerr^2+n^2}\,.$ 
The coordinate change 
\begin{align}\label{eq:kerr8}
r & =r_+ + \frac{r_+-r_-}{4\Upgamma_+}\,\uprho^2\,, \qquad 
\phi =\varphi+\frac{2{\kappa} a}{r_{+}-r_{-}}{t},\nonumber
\end{align}
with $\Upgamma_+ =r_+^2+(n+a\cos\theta)^2$ 
shifts the Kerr black hole horizon at $r=r_+$ to $\uprho=0$ and yields the metric into a suitable form for the near horizon expansion, given by \cite{daniel3}
\begin{subequations}
\label{eq:kerreqs}
\begin{align}
g_{tt} & = -\kappa_{\text{\tiny{H}}}^2 \uprho^2 +{\cal O}\left(\uprho^{4}\right),\qquad g_{\rho\rho} =1+{\cal O}\left(\uprho^{2}\right),\qquad g_{\theta\theta} =\Upgamma_++O\left(\uprho^{2}\right),\nonumber \\
g_{t\varphi} &= \frac{\kappa_{\text{\tiny{H}}}}{2\Upgamma_{+}^{2}}\left[2ar_{+}\sin^{2}\theta\left(2n^{2}+r_{+}^{2}+r_{-}r_{+}\right)
+\left(r_{+}-r_{-}\right)\left(a\sin^{2}\theta-2n\cos\theta\right)\Upgamma_{+}\right]\uprho^{2}+{\cal O}\left(\uprho^{4}\right)\nonumber
\\
g_{\rho\theta} & =\frac{a (n+a\cos\theta)\sin\theta}{\Upgamma_+}\,\uprho+{\cal O}\left(\uprho^{3}\right),\qquad g_{\varphi\varphi} =\frac{\left(2n^2+r_+(r_++r_-),\nonumber \right)^{2}\sin^{2}\theta }{\Upgamma_+} + {\cal O}\left(\uprho^{2}\right),\nonumber\\
g_{t\rho}& =g_{t\theta}=0=g_{\rho\varphi} =g_{\theta\varphi}.
\nonumber
\end{align}
\end{subequations} 
In particular, the horizon metric reads 
\begin{equation}
\Omega_{ij}\,\mathrm{d} x^i\mathrm{d} x^j = \Upgamma_+\,\mathrm{d}\theta^2 
+\frac{1}{\Upgamma_+}\left(2n^2+r_+(r_++r_-) \right)^{2}\sin^{2}\theta\,\mathrm{d}\varphi^2\,.\nonumber
\label{eq:kerr6}
\end{equation}
\bibliography{bibliografia}
\bibliographystyle{iopart-num}

\end{document}